# Manipulation of ferroelectric and magnetic properties properties of single phase $Bi_4Ti_3O_{12}$-$3BiFeO_3$ solid solution through La substitution

K.R.S. Preethi Meher* and K.B.R.Varma

*Materials Research Centre, Indian Institute of Science, Bangalore 560012*

*Corresponding authors Email: preethimeher@cutn.ac.in and kbrvarma@mrc.iisc.ernet.in

*author presently affiliated to

*Department of Materials Science*
*School of Technology*
*Central University of Tamilnadu*
*Neelakudy campus*
*Thiruvaur 610005*

## Abstract

In this work, the potential to exploit Aurivillius higher order composition (n=6) given by $Bi_4Ti_3O_{12}$-$3BiFeO_3$ as single phase and high temperature magnetoelectric materials are well explored by elaborately studying the concerned structure and physical properties. $Bi_4Ti_3O_{12}$-$3Bi_{1-x}La_xFeO_3$ (x=0, 0.01, 0.05, 0.1, 0.2) was synthesized via solid-state reaction route and X-ray diffraction data revealed that the lattice parameters associated with orthorhombic cell (space group $P2_1am$) decrease with increase in $La^{3+}$ content confirming the incorporation of La. In the frequency dependent dielectric measurements, the ferroelectric to paraelectric transition temperature ($T_c$) corresponding to the parent (x=0) had two phase transitions vis., 803 K (broader) and 983 K (sharper) whereas phase transitions are single that progressively becomes sharper with increase in La substitution. Ferroelectric switching behavior (polarization (P) versus electric field (E)) was demonstrated for $Bi_4Ti_3O_{12}$-$3BiFeO_3$ and remnant polarization decreased significantly for La substituted compositions reflecting weaker ferroelectric ordering. In detailed Zero Field-cooled and Field cooled temperature dependent magnetization measurements, $Bi_4Ti_3O_{12}$-$3BiFeO_3$ exhibited $T_c$ around 296 K. Though a marginal increase in the magnetization values for La substituted compositions (x=0.1 and x=0.2), the overall temperature dependent magnetization curves indicate overall paramagnetic nature. The results are more reflective of the switching off of the possibilities of long range ferromagnetic ordering with A-site rare earth doping in these Aurivillius layered compounds.

# 1. Introduction

In the last decade, $BiFeO_3$, a perovskite belonging to R-3*m* space group, has attracted the attention of many active research groups due to its ferroelectric ordering below 1100 K and a G-type non-collinear antiferromagnetic ordering below 643 K [1-5]. However resistivity properties of $BiFeO_3$ had to be improved vis-à-vis to minimize its dielectric losses. Solid solutions of $BiFeO_3$ with other perovskite $ABO_3$ type ferroelectric materials like $BaTiO_3$, $PbTiO_3$, $Na_{0.5}Bi_{0.5}TiO_3$, $PrFeO_3$, $PbNb_{0.5}Fe_{0.5}O_3$, $Bi_4Ti_3O_{12}$ etc. are being studied in a bid to better its ferroelectric properties accompanied by low leakage currents [4-21]. The $Bi_4Ti_3O_{12}$-$BiFeO_3$ solid solution system giving rise to a series of compounds with the formula $Bi_2Bi_{n-1}(Ti,Fe)_nO_{3n+3}$ where n is 3,4,5 etc. also has been investigated for its ferroelectric and magnetoelectric properties since 1967 [6]. They belong to the Aurivillius family with general formula $(Bi_2O_2)^{2+}$ $(A_{n-1}B_nO_{3n+1})^{2-}$ where the fluorite like bismuth oxygen layers of composition $(Bi_2O_2)^{2+}$ is stacked alternatively with the (001) oriented perovskite blocks of composition $(A_{n-1}B_nO_{3n+1})^{2-}$. Cations such as $Bi^{3+}$, $Ba^{2+}$, $Ca^{2+}$, $Sr^{2+}$, $La^{3+}$ occupy the cubo-octahedral A-site in the perovskite slab and the octahedral B-sites within the perovskite slab are occupied by $W^{6+}$, $Ti^{4+}$, $Nb^{5+}$, $Fe^{3+}$, $Ta^{5+}$ etc. and n represents the total number of perovskite layers within the unit cell [7]. The tolerance factor of these compounds is in the range of 0.87 – 0.97, which is favorable for the formation of layered structures with a desired number of perovskite layers within a unit cell. As $Bi_4Ti_3O_{12}$ is an n=3 Aurivillius phase, a solid solution consisting of $Bi_4Ti_3O_{12}$ and $mBiFeO_3$ (where m=1, 2, 3 etc.) also results in Aurivillius phase with higher n (where n is 4, 5, 6, 7 etc.) [21-36]. Mostly the focus has been on the possibilities to exploit their potential multiferroic properties by exploring the possible compositions in both polycrystalline and thin film forms. When a paramagnetic transition metal ion such as Fe, Mn, Co, Ni or Cr can occupy the B-site of the Aurivillus compound without affecting its crystal symmetry, there is a huge possibility to end up in a phase which can be simultaneously ferroelectric and ferromagnetic/antiferromagnetic/ ferrimagnetic resulting in multiferroic behavior [37-46]. However this is quite a challenge as $3d^0$–ness of the B-site cations is essential to maintain the ferroelectric ordering [9]. Interestingly, Mao et al reported above - room temperature ferromagnetic and ferroelectric ordering of Co substituted $Bi_5Ti_3FeO_{15}$ [38]. Though a number of reports on the magnetic properties of $Bi_4Ti_3O_{12}$-n$BiFeO_3$ system where n ≤ 5 are already available in the literature, detailed magnetic studies of n>5 Aurivillius compounds are few and incomplete. Infact the structure and ferroelectric properties of $Bi_4Ti_3O_{12}$-5 $BiFeO_3$ (BFT5) system has been systematically studied earlier [9,10]. Only recently, on using advanced structural characterization techniques such as Transmission electron microscopy (TEM) and neutron diffraction studies, it has been found that the Aurivillius compounds with higher n is prone to be present as intergrowths and riddled with stacking faults and other defects. In fact, the present authors have discussed the intergrowth nature of $Bi_4Ti_3O_{12}$-5$BiFeO_3$ backed by detailed TEM studies [32]. The present work focuses on $Bi_4Ti_3O_{12}$-3$BiFeO_3$ (m=3; n=6) (BFTO) which possesses certain promising attributes of being a good magnetoelectric material. Dielectric studies carried out previously on this system have been largely influenced by high ionic conductivity generally associated with such ceramic systems [11, 21]. For the same reason, its ferroelectric switching behavior has not been satisfactorily demonstrated so far. Over the course of different substitution related studies, A-site substitution of $La^{2+}$ in some of the Aurivllius phases [34-36] has been found to improve its overall properties. Hence an investigation into the effect of La substitution in $Bi_7Ti_3Fe_3O_{21}$ may deserve attention. In this article, we report the details concerning the structural, dielectric, ferroelectric and magnetic properties of $Bi_4Ti_3O_{12}$-$3BiFeO_3$ (BFTO), $Bi_4Ti_3O_{12}$-$3La_xBi_{1-x}FeO_3$ (x=0.01, 0.05, 0.1, 0.2) ceramics fabricated using the powders synthesized by solid-state reaction route.

## 2. Experimental

The $Bi_4Ti_3O_{12}$-$3La_xBi_{1-x}FeO_3$ (x=0.0, 0.01, 0.02, 0.1 and 0.2) polycrystalline powders were synthesized by conventional solid-state reaction method. Stoichiometric amounts of $La_2O_3$, $Bi_2O_3$, $Fe_2O_3$ and $TiO_2$ that are required to obtain $Bi_4Ti_3O_{12}$-$3La_xBi_{1-x}FeO_3$ were mixed together by ball milling in acetone medium for 24 h and the milled powders were calcined at various temperatures (973 K-1273 K) till the desired Aurivillius phase was obtained. A Bruker D8 Advance with in-situ high temperature X-ray powder diffractometer (XRD) with a $CuK_\alpha$ source was employed to obtain the details with regard to the phase transition in BFTO and the $La^{3+}$ substituted samples. The room temperature XRD data were collected at a scan speed of $0.05^o$/min and a step size of $0.048^o$ in the 2θ range of $15^o – 80^o$. The high temperature XRD data were collected over a narrow 2θ range ($28^o – 33^o$) that encompass the high intensity (1 1 13) and (200) reflections in steps of 50 K from 473 – 1023 K temperature range. Profile refinements of the XRD patterns were carried out using TOPAS3.0 in order to index the peaks and calculate the lattice parameters. The dielectric properties of BFTO pellets (sintered at 1098 K/24h) sputtered with gold electrodes on both the sides were studied in a wide frequency range (100 Hz to 10 MHz) at various temperatures (300 – 1023 K) using HP4194 impedance analyzer. The ferroelectric hysteresis loops were recorded using a loop tracer based on Sawyer-Tower circuit. DC magnetization (M) measurements including the Magnetization switching behavior (M Vs H) were carried out using a Super conducting Quantum interference device, Vibrating Sample magnetometer with cyrogenic set up between 10 K to 300 K temperature range at an applied magnetic field of 500 Oe to 1 kOe under both Zero-Field cooled (ZFC) and Field-cooled (FC) conditions. Magnetization measurements under AC conditions were carried out using AC susceptometer under an applied AC magnetic field of 0.15 Oe. Scanning Electron Microgrph images recorded for sintered specimens has been given in the Supporting information Fig. S1.

## 3. Results and discussion

### 3.1 XRD studies

**Fig.1** shows the XRD patterns obtained at room temperature for BFTO and that of the various amounts of $La^{3+}$ substituted BFTO (x=0, 0.05, 0.1 and 0.15) and the Bragg peaks were indexed according to that of an n=6 Aurivillius phase (according to structural details in Ref[10]). Decomposition to multiple phases occurred when we tried to synthesize compositions containing higher substitutions of La (x>0.2). From the profile refinement of XRD patterns using TOPAS 3.2 shown in **Fig. 2(a)**, the lattice parameters of BFTO were found to be a=5.523 Å; b=5.614 Å and c= 58.651 Å based on the orthorhombic space group $P2_1am$ ($R_{wp}$ ~ 6.2 %). A similar procedure was carried out to obtain the lattice parameters for all the La substituted compositions (as listed in **Table 1**). The profile refinement of the XRD data recorded for LBFTO4 alone has been shown in **Fig. 2(b)** for representation. The lattice parameters were found to decrease overall with increase in La substitution – confirming the systematic incorporation of La within the perovskite layers. Traces of secondary phase viz $Bi_2Fe_4O_9$ [25], was found to be present along with BFTO and the phase becomes too little to be observed for La substituted compositions.

**3.2 High temperature XRD studies**

The high temperature XRD studies were carried out for BFTO ceramics in order to ascertain the nature of the structural transition. The evolution of XRD peaks corresponding to (200) (consisting of closer (200) and (020) reflections) of BFTO was closely monitored for different temperatures from 473 K to 1023 K as shown in **Fig. 3(a)**. As the (11 13) reflection only concerns with change in the lattice parameter "c", the corresponding temperature dependent data has not been considered here for analysis. The full width at half maxima (FWHM) of (200) (as shown in **Fig. 3(b)**) was calculated by fitting them with a Lorentzian function. A significant reduction in FWHM was observed around 900 K, where a isostructural phase transition (both being orthorhombic) occurs during which the two closer peaks corresponding to (200) and (020) coalesce leading to a single peak. In order to have a good estimate of the structural transition temperature, the lattice parameters (a,b) were calculated from the 2θ positions of (200) reflections and were plotted as a function of temperature **Fig. 3(c).** The incidence of an anomaly in the temperature dependent lattice parameter curves ( a and b) is evident around 925 K. Above 925 K, the difference between a and b lattice parameters starts to diminish drastically but does not become zero indicating that the orthorhombic to tetragonal phase transition never happens even at these higher temperatures – reflective of high orthorhombic structure stability. Very high transition temperatures have been a characteristic feature of Aurivillius phases of n > 4 [31]. For compositions substituted with $La^{3+}$, the decline in FWHM values of (200) reflection, whch is a sign of the phase transition (shown in Fig. 4 (a-d)), shifts toward lower temperatures ~700 K with increase in La level.

## 4. Dielectric studies

The dielectric behavior of the sintered samples of BFTO and La substituted BFTO were studied in the 300 K – 1000 K temperature range at different frequencies (100 Hz to 10 MHz). **Fig. 4** shows the variation of dielectric constant ($\varepsilon'_r$) and loss (D) as a function of temperature at a representative frequency of 500 kHz. At this frequency, $\varepsilon'_r$ is ~ 80 and the corresponding D is ~ 0.009 at room temperature. Two dielectric anomalies were observed in the temperature dependent dielectric behavior viz., 803 K and 983 K respectively. The broad diffusive peak around $T_M$=803 K was observed for all the frequencies under investigation. The ferroelectric to paraelectric phase transition was observed around 983 K which is accompanied by a peak around 974 K in the dielectric loss (D) curve **(marked in Fig. 4)**. The transition is not very prominent at low frequencies (<100 kHz) due to finite conductivity associated with this system at higher temperatures. The broad and diffuse peak observed around 803 K in the case of BFTO may or may not involve a structural transition as this is not very clear from the high temperature XRD studies discussed above. The cooperative relaxation of defect dipoles associated with the oxygen vacancies could be a possible reason for the appearance of diffuse peak in the dielectric behavior. The temperature dependent dielectric constant ($\varepsilon_r'$) and loss (D) of La substituted BFTO samples at certain selected frequencies is shown in **Figs. 5(a-d) and 6(a-d)**. One could observe a variation in the overall dielectric behavior for the La substituted samples ranging from relaxor to the conventional ferroelectric-like characteristics. $\varepsilon_r'$ variation with temperature for x=0.05 is highly diffusive with no clear features of a sharp ferroelectric to paraelectric phase transition. However broader peaks that shifted to higher temperatures with increasing frequency were observed in the temperature dependent loss (D) behavior (**Fig. 6 (a)**). The temperature $T_M$ (where D exhibits a peak) at 100 Hz is ~ 761 K. There are no strong relaxor-like features in the temperature dependent dielectric loss (D) behavior of x=0.1 with only a very weak dispersion in the peaks as well as reduced diffusivity. Here, the maximum in the dielectric loss occurs at ~ 860 K for 100 Hz of applied frequency. For x=0.15 and x=0.2, much sharper anomalies were observed in both the temperature dependent $\varepsilon_r'$ and D behavior implying a conventional ferroelectric to paraelectric phase transition with $T_c$ values ~760 K and 749 K respectively. The values of the phase transition temperatures ($T_c$) along with the dielectric constant ($\varepsilon_r'$) and loss values of BFTO and the La substituted BFTO have been listed in **Table 2**. $T_M$ value recorded at an applied frequency of 100 kHz has been reported (**in Table 2**) instead of $T_c$ for x=0.05 owing to its diffusive nature. There is a significant decrease in the ferroelectric to paraelectric transition temperature values when Bi is partially substituted by La in BFTO. The log$\omega_{max}$ (where $\omega_{max}$ is the frequency corresponding to the maximum in $\varepsilon_r'$) versus the $T_m$ plot has been given in Fig. S2 (Supporting information). This experimental data were fit with the mathematical empirical model known as Voger-Fulcher (V-F) equation given by $\omega=\omega_0 \exp(U/k(T-T_f))$ where the $\omega_0$ is the average relaxation time, U is the activation energy and $T_f$ is the freezing temperature. The best fit (solid line in **Fig. S2**) to the experimental data according to the V-F equation was obtained. The freezing temperature was estimated to be around 691.99 ± 20.46 K with an average relaxation frequency ($\omega_0$) of 1.599X10$^7$ Hz and activation energy ($E_a$) value of 0.061 eV. On the whole, when the RT/high temperature structural and dielectric studies are put together, one can understand that the ferrodistortive modes are weakened at relative lower temperatures on La substitution as conveyed by the lowered $T_c$ values compared to BFTO phase. However, one can see that the phase transitions becomes sharper and definite for high substitution levels of La, devoid of any diffusive relaxation behavior (like those observed for BFTO) which is indicative of the pinning of oxygen vacancies.

## 5. Ferroelectric properties

Though structural and dielectric studies of BFTO and x=0.2 samples provide enough evidence for the ferroelectric properties of these systems, study of the polarization (P)-electric field (E) behavior would be necessary to understand its ferroelectric nature. Ferroelectric hysteresis loops of BFTO were obtained at an applied frequency of 50 Hz around 473 K. Remnant polarization ($2P_r$) value of 8.9 $\mu C/cm^2$ associated with the corresponding coercive field ($E_c$) of 40 kV/cm was observed in the case of BFTO (shown in **Fig. 8 (a)**). The maximum polarization value ($P_{max}$) of 6.8 $\mu C/cm^2$ could be obtained at a maximum applied field of 65kV/cm. Though P-E hysteresis behavior is suggestive of ferroelectric nature, saturated rectangular loops could not be obtained as the application of higher electric field was limited due to its poor resistive properties. In the case of La substituted composition, the P-E hysteresis behavior indicates the weakening of the ferroelectric property. The evolution of hyteresis loops with increasing applied electric field is shown in the **Fig. 8 (b)**. Very low $P_r$ values ~ 0.6 $\mu C/cm^2$ were obtained even at a maximum applied field of 75 kV/cm. These results are important as they show that it is possible to get well-defined hysteresis loops for BFTO. As the ionic radii of $La^{3+}$ is very close to that of $Bi^{3+}$ ($R_{La}^{3+}$=1.16 Å, $R_{Bi}^{3+}$=1.17 Å), the decrease of $T_c$ may not be attributed only to the change in tolerance factor. The replacement of highly polarizable $Bi^{3+}$ with less polarizable $La^{3+}$ cations within the perovskite layers seems to be a yet another important factor contributing to the decrease in $T_c$. This is corroborated by the weakened orthorhombicity (a/b increases from ~ 0.98 to 0.99) in the case of different La compositions as reflected in their calculated lattice parameters values reported in **Table 1**. Very low polarization values (~ 0.5 $\mu C/cm^2$) and unsaturated P – E loops indicate only a partial switching of ferroelectric domains unlike in the case of pure BFTO.

**Magnetic studies**

The DC magnetization behavior of BFTO was studied by subjecting the sample to ZFC and FC conditions under an applied magnetic field of 500 Oe from 10 K to 325 K as shown in Fig.9(a). One can observe a weak anomaly around 250 K in both the ZFC and FC curves though the overall trend in the magnetization, especially at temperatures less than 100 K mimics a typical paramagnetic system. However, if one observes carefully, the ZFC and the FC curves coalesce together at around 296 K that indicates the presence of a possibly a weak ferri or antiferromagnetic transition around that temperature. As the XRD analysis of BFTO has revealed the presence of $Bi_2Fe_4O_9$ in trace amounts (less than 1%), the weak anomaly around 250 K can be attributed to its antiferromagnetic to paramagnetic transition [45]. In order to confirm the presence of transition around 296 K, the BFT3 was subjected to AC susceptibility measurement from room temperature down to 100 K at a very small applied ac field of 0.17 Oe and 420 Hz. We employed this technique as both the in-phase ($\chi'$) and the out of phase ($\chi''$) components of AC susceptibility are very sensitive to the weak thermodynamic phase changes. The inset in Fig. 9(a) shows the variation of $\chi'$ with temperature where an anomaly is observed near 300 K in a way confirming the DC magnetization results. Magnetic hysteresis loops (M Vs H) behavior has been recorded for BFTO at 300 K and 5 K at an applied field of -3 T to 3 T (shown in Fig. 9(b)). The room temperature M-H behavior (recorded at 300 K), though looks linear overall, exhibits a non-linear behavior with extremely small remnant magnetization of around ~2.9 memu/g and coercivity value of 184 Oe (the zoomed image of the M-H data at 300 K is shown in the inset of Fig. 9 (b)). At 5 K, the remnant magnetization increases five times of that observed at room temperature thought the overall trend remains the same. Earlier studies by Dong et al reported a similar M-H curves for bulk $Bi_4Ti_3FeO_{15}$ where it was interpreted as a superparamagnetic behavior [40]. However the superparamagnetism occurs to due to a net giant spin (develops when particle size is comparable to domain size), that switches without hysteresis below a blocking temperature. Such an effect has a less probability to occur in the present systems which mostly contains only short range ordered clusters (though likely to be less than 100 Å) that very weakly interacts with each other. The magnetic measurements were also carried out on La doped BFTO samples (x=0.1 and x=0.2 were considered for studies) from 325 K to down to 10 K at an applied field of 500 Oe under both the ZFC and FC conditions (Fig. 10 (a) and (b)). Contrary to the expectation that the La doping might strengthen the ferromagnetic interaction in the sample, we obtain an overall paramagnetic trend in the magnetization curves for both the systems. The inverse susceptibility curve ($1/\chi$ Vs T shown as inset of Fig. 10(a)) was plotted for x=0.1 and the high temperature region (> 280 K) was fit with Curie-Weiss equation given by $\frac{1}{\chi} = \frac{(T-\theta)}{C}$ where C is the Curie-Weiss constant and θ is the transition temperature. The calculated effective magnetic moment '$p_{eff}$' ($g\sqrt{s(s+1)}$) was $4.54 \pm 0.04\mu_B$ which is significantly lower than the theoretically expected value of 5.91 for $Fe^{3+}$ cation. The 'θ' value was calculated to be -112 ± 2K which indicates the presence of very weak antiferromagnetic interactions in the system. It looks like the La doping actually switches off the active ferromagnetic interactions that we found in the parent Aurivillius phase instead of promoting them.

**6. Conclusions**

Ferroelectric to paraelectric phase transition around 983 K was experimentally confirmed for $Bi_4Ti_3O_{12}$-$3BiFeO_3$. The remnant polarization value of $2P_r \sim 8.9$ μC/cm$^2$ with the corresponding coercive field of 40 kV/cm were obtained whereas the La substituted samples, whose $T_c$ values were found to shift to lower temperatures (< 900 K), showed a weak ferroelectric switching behavior as demonstrated for the x=0.2 composition. The magnetic properties of $Bi_4Ti_3O_{12}$ - $3BiFeO_3$ along with La doped compositions (x=0.1 and x=0.2) were studied in detail. $Bi_4Ti_3O_{12}$ - $3BiFeO_3$ exhibited a ferromagnetic to paramagnetic phase transition near RT (around 296 K) and the La substituted compositions were more paramagnetic in nature. However the magnetization behavior for the parent Aurivillius phase could be confirmed by adopting different synthetic approaches where the traces of secondary phase can be completely avoided.

*Table 1* List of the values of lattice parameters obtained from profile refinement for La substituted BFTO

| Composition (x) | a (Å) | b (Å) | c (Å) |
|---|---|---|---|
| 0.0 | 5.523 (7) | 5.614 (8) | 58.65 (12) |
| 0.01 | 5.489 (3) | 5.523 (4) | 57.551 (3) |
| 0.05 | 5.504 (7) | 5.559 (8) | 58.027 (7) |
| 0.1 | 5.483 (7) | 5.526 (6) | 57.566 (11) |
| 0.2 | 5.471 (8) | 5.521 (8) | 57.468 (11) |

*Table 2* List of transition temperature ($T_c$), relative dielectric constant ($\varepsilon'_r$) and dissipation factor (D) at room temperature for the La substituted BFTO

| Compn. (x) | $T_c$ (K) | $\varepsilon'_r$ (300 K & 500 kHz) | D (300 K & 500 kHz) |
|---|---|---|---|
| 0 | 983 | 75 | 0.009 |
| 0.01 | 908 | 92 | 0.099 |
| 0.05 | 893 | 128 | 0.079 |
| 0.1 | 760 | 103 | 0.067 |
| 0.2 | 749 | 88 | 0.04 |

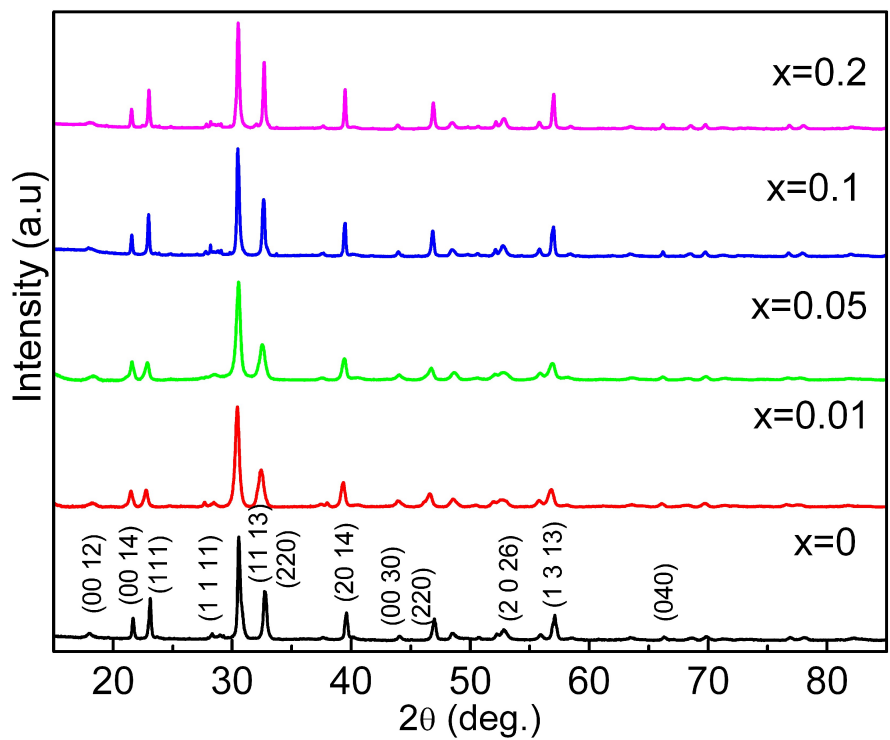

**Fig. 1** X-ray diffraction patterns of (a) BFTO and La doped BFTO (x=0, 0.01, 0.05, 0.1 and 0.2).

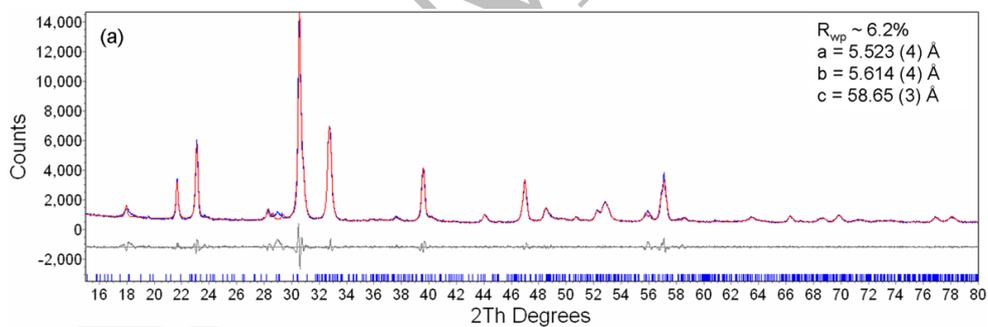

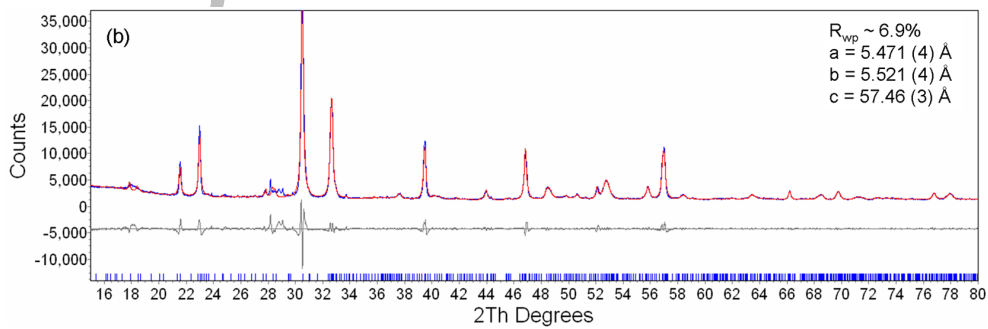

**Fig. 2 Profile refinement of** X-ray diffraction patterns of (a) BFTO and (b) La doped BFTO (x=0.2). The experimental (red curve) and the calculated (blue curve) diffraction data are shown along with the difference (grey curve) and the allowed Bragg reflections.

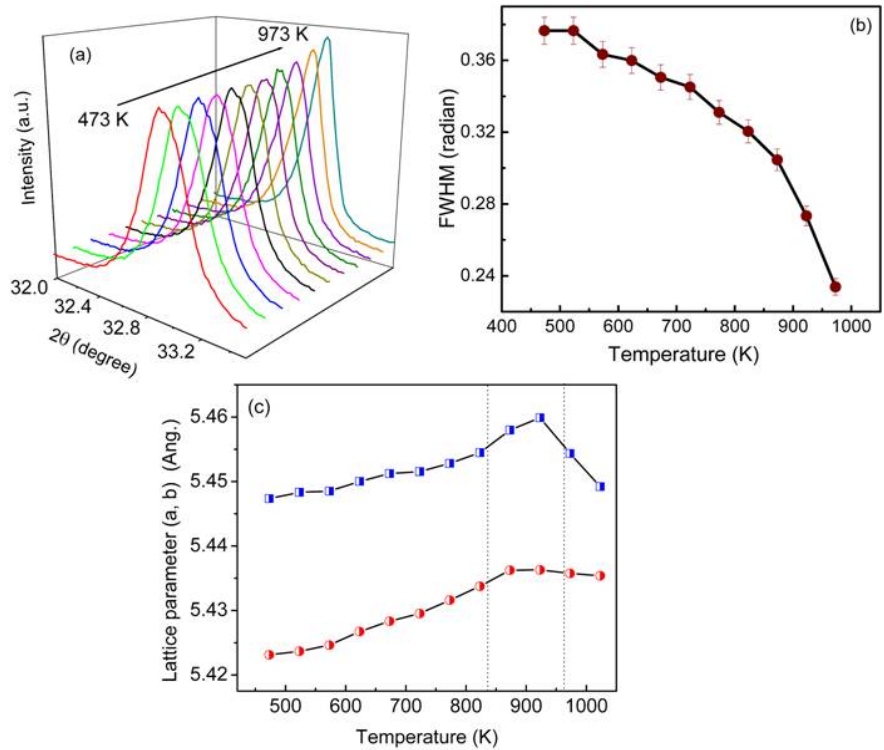

**Fig. 3** High temperature X-ray diffraction data recorded for BFTO: (a) evolution of (200) peak from 473 K to 973 K (b) FWHM of the (200) peak and (c) change in calculated lattice parameters (a,b) as a function of temperature indicating an isostructural phase transition

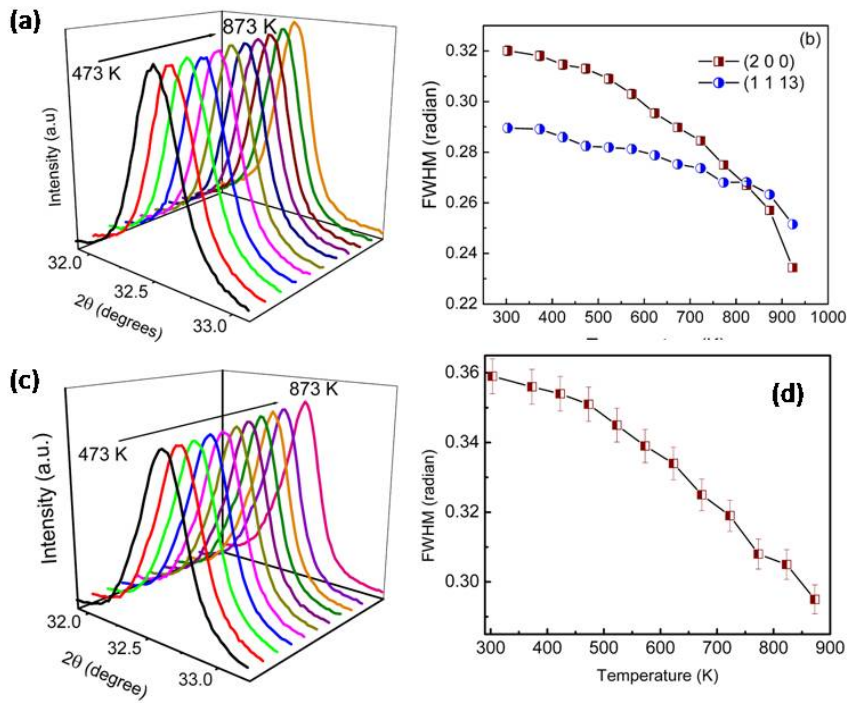

**Fig. 4** High temperature X-ray diffraction data recorded for x=0.1 and x=0.2 compositions: (a) evolution of (200) peak from 473 K to 973 K for x=0.1 (b) FWHM of the (200) and (11 13) peak for x=0.1 (c) evolution of (200) peak from 473 K to 973 K for x=0.2 (d) FWHM of the (200) peak for x=0.2

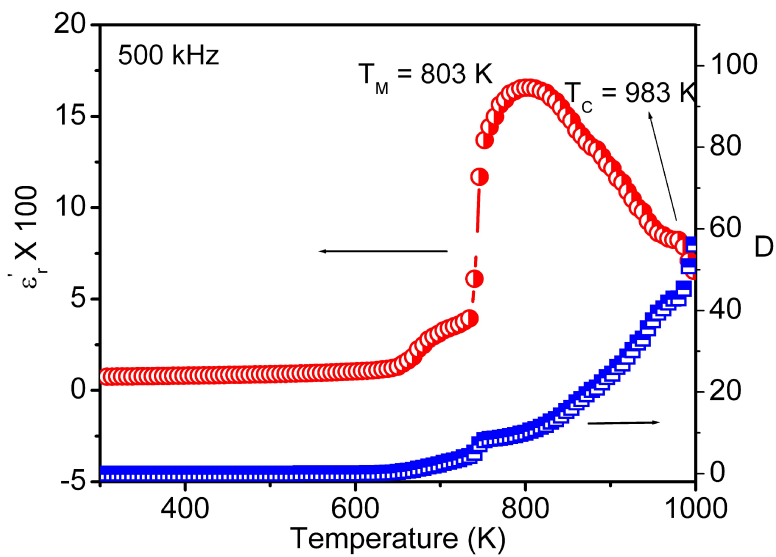

**Fig.5** Variation of relative dielectric constant ($\varepsilon_r'$) and loss (D) as a function of temperature (300 K – 1000 K) at 500 kHz for BFTO

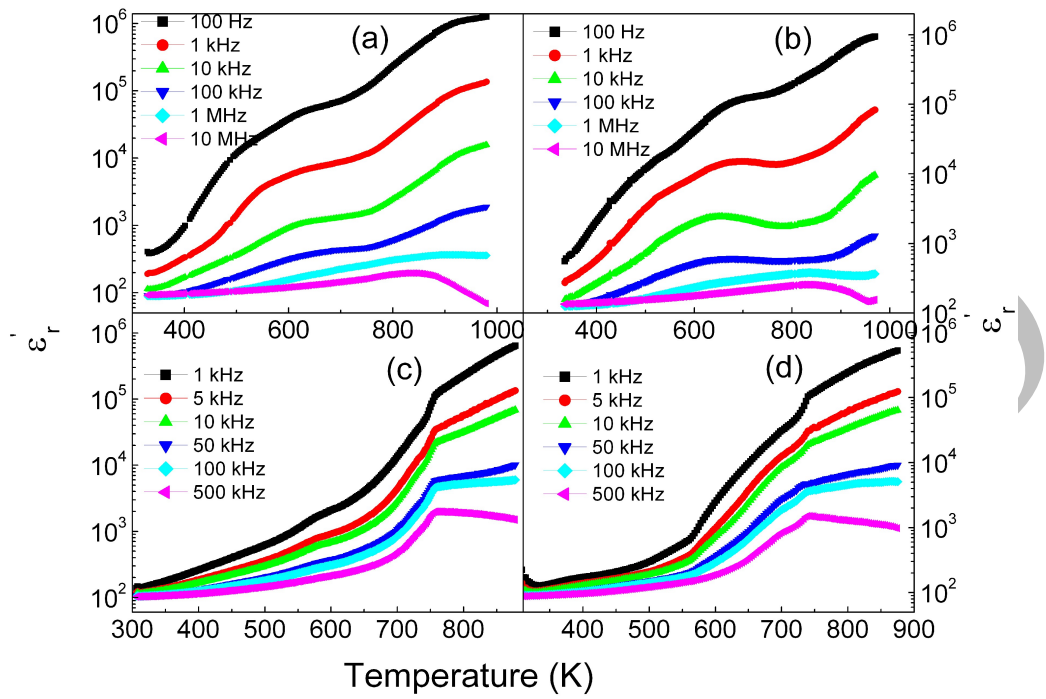

**Fig. 6** Variation of dielectric loss factor (D) as a function of temperature for (a) x=0.05 and (b) x=0.1 (c) x=0.15 and (d) x=0.2 respectively.

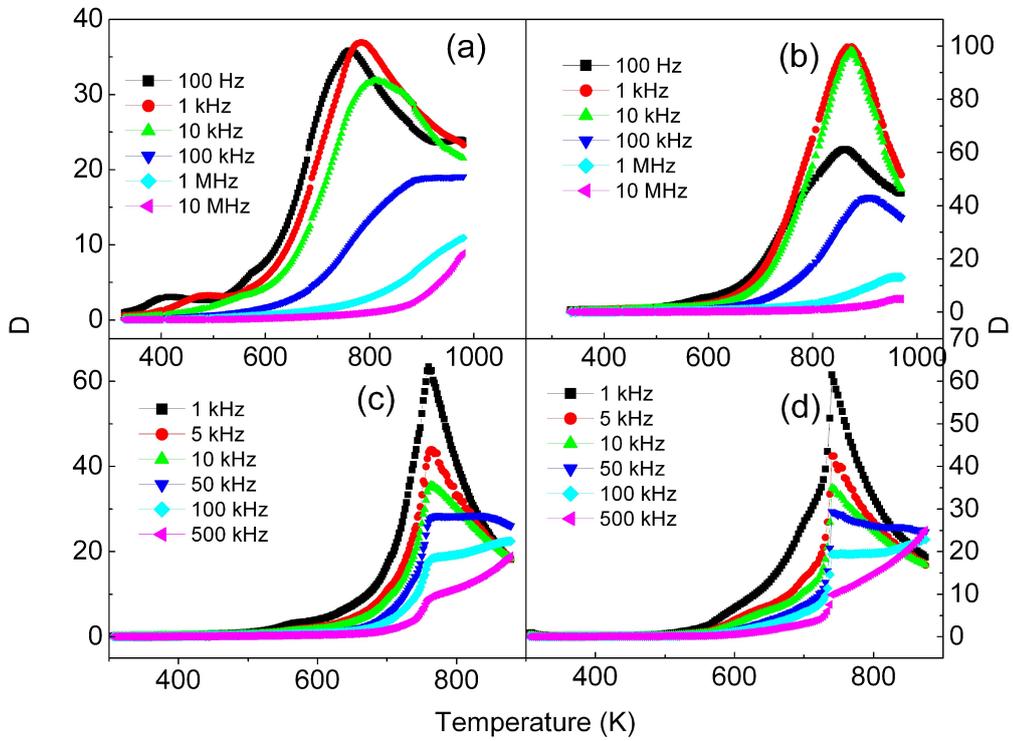

**Fig. 7** Variation of dielectric loss factor (D) as a function of temperature (a) x=0.05 (b) x=0.1 (c) x=0.15 and (d) x=0.2 respectively

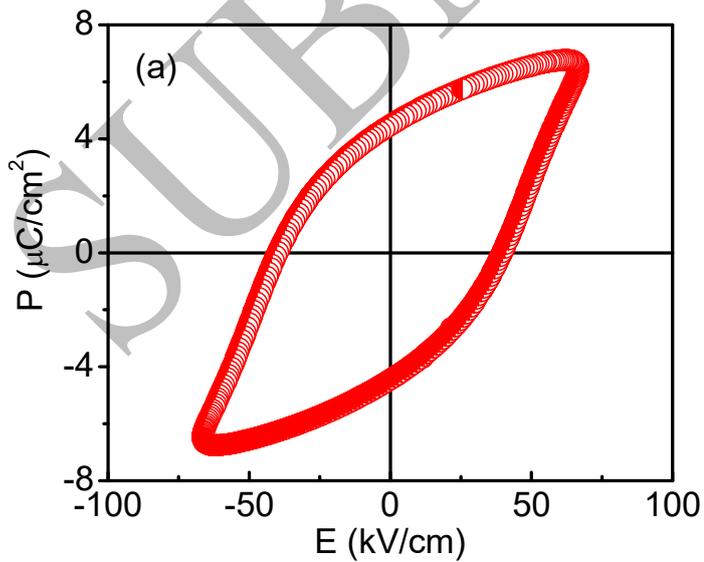

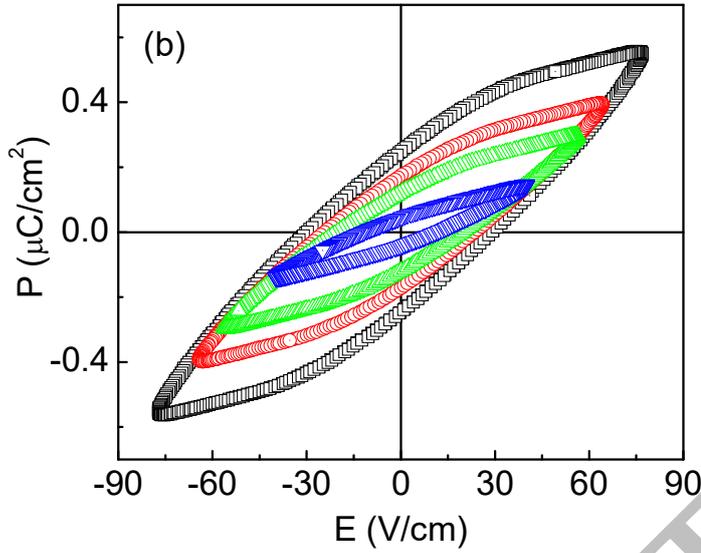

**Fig.8** Ferroelectric P-E hysteresis loop of (a) BFTO at 450 K and (b) x=0.1 with the evolution of the hysteresis loop with increasing electric field values

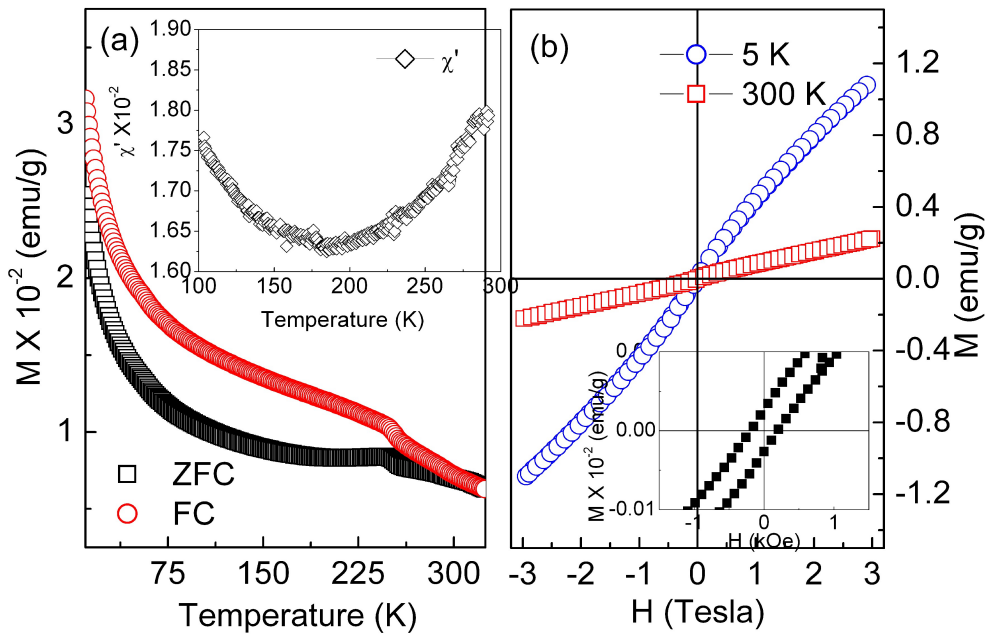

**Fig.9** Variation of magnetization with temperature from 5 K to 325 K under both the ZFC and FC conditions at an applied field of 500 Oe for BFTO. The **inset** shows the temperature dependence of AC susceptibility ($\chi'$) from 100 to 300 K at an AC field of 0.17 Oe and 420 Hz **(b)** Variation of magnetization (M) with applied magnetic field (H) of ± 3 Tesla at 300 K and 5 K. Inset is the zoom of M-H curve from -1T to 1T

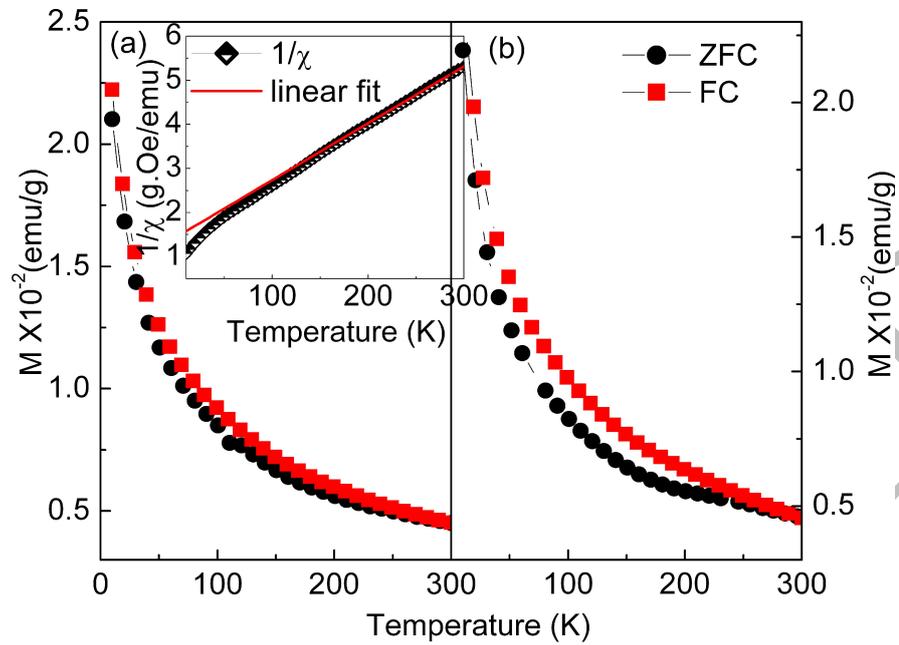

**Fig. 10** Variation of magnetization with temperature from 10 K to 300 K under both the ZFC and FC conditions at an applied field of 500 Oe for **(a)** x=0.1 and **(b)** x=0.2. The **inset of Fig.10(a)** is the variation of inverse susceptibility with Temperature (K) for x=0.1 composition and the solid line is the fit according to Curie-Weiss law.

Supplementary information for "**Manipulation of ferroelectric and magnetic properties properties of single phase Bi$_4$Ti$_3$O$_{12}$-3BiFeO$_3$ solid solution through La substitution**"

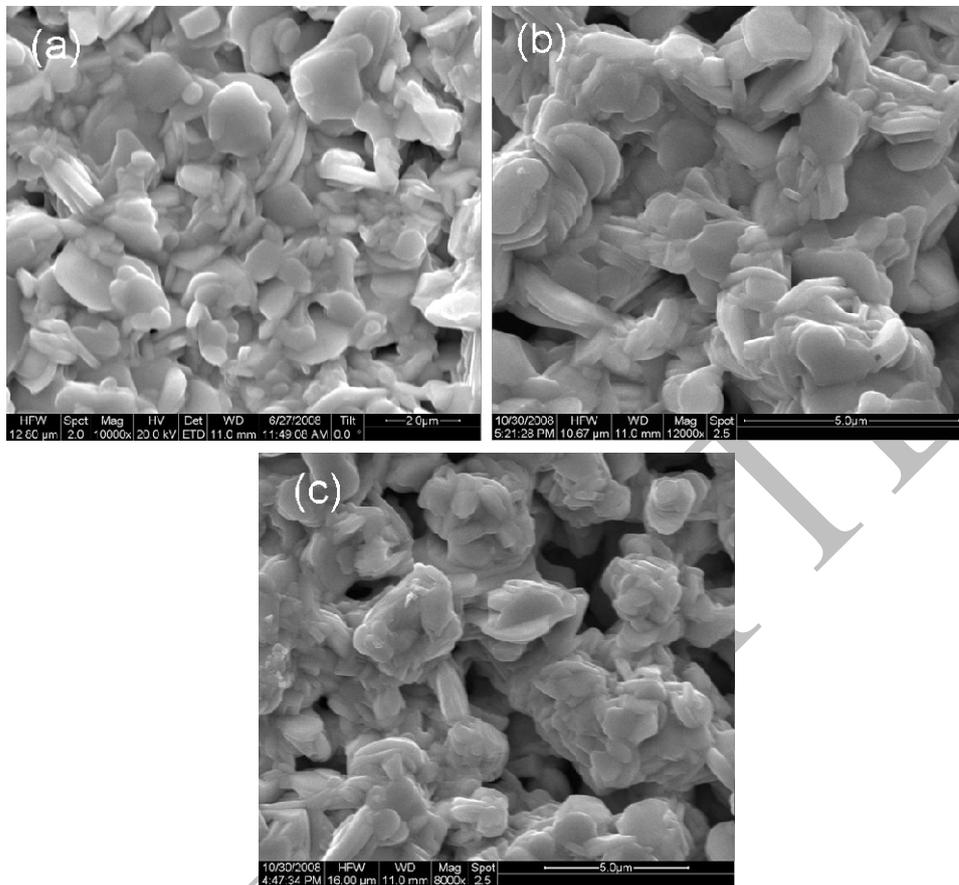

**Fig. S1** SEM images of the sintered samples corresponding to (a) BFTO (b) x=0.1 and (c) x=0.2 respectively

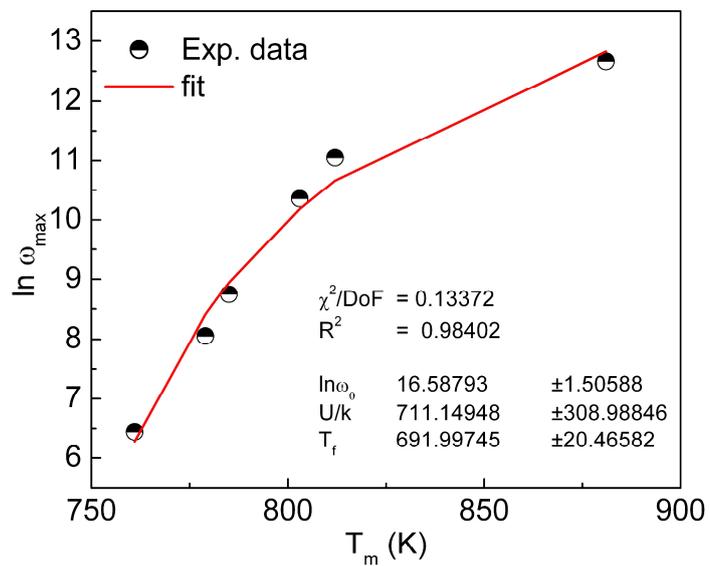

**Fig. S2:** $\ln\omega_{max}$ as a function of peak temperature ($T_M$) for x and Vogel-Fulcher fit of the experimental data.